\documentclass[aps,prl,floats,preprintnumbers,showpacs,twocolumn]{revtex4}
\usepackage{epsfig}

\begin{document}

\preprint{NUHEP-TH/05-02}

\title{See-Saw Energy Scale and the LSND Anomaly}

\author{Andr\'e de Gouv\^ea}
\affiliation{Northwestern University, Department of Physics \& Astronomy, 2145 Sheridan Road, Evanston, IL~60208, USA}


\begin{abstract}

The most general, renormalizable Lagrangian that includes massive neutrinos contains ``right-handed neutrino'' Majorana masses of order $M$. While there are prejudices in favor of $M\gg M_{\rm weak}$, virtually nothing is known about the magnitude of $M$. I argue that the LSND anomaly provides, currently, the only experimental hint: $M\sim 1$~eV. If this is the case, the LSND mixing angles are functions of the active neutrino masses and mixing and, remarkably, adequate fits to all data can be naturally obtained. I also discuss consequences of this ``eV-seesaw'' for supernova neutrino oscillations, tritium beta-decay, neutrinoless double-beta decay, and cosmology.

\end{abstract}

\maketitle

Neutrino experiments have revealed, beyond reasonable doubt, that neutrinos have mass and mix. Neutrino masses indicate the existence of physics beyond the standard model (SM), and we do not yet have enough information to determine what is this new physics \cite{TASI}.

A very simple mechanism for explaining neutrino masses is the following. Add to the old SM Lagrangian gauge singlet Weyl fermions $N_i$, $i=1,2,3$ \cite{foot1}, and allow for the most general renormalizable Lagrangian consistent with gauge invariance. Under these circumstances, the ``new standard model'' Lagrangian is
\begin{equation}
\label{eq:seesaw}
{\cal L}_{\nu}={\cal L}_{\rm old} - \lambda_{\alpha i}L^{\alpha}HN^i-\sum_{i=1}^3\frac{M_i}{2}N^iN^i + H.c.,
\end{equation}
where $H$ is the Higgs weak doublet, $L^{\alpha}$,  $\alpha=e,\mu,\tau$, are lepton weak doublets, $\lambda_{\alpha i}$ are neutrino Yukawa couplings, and $M_i$ are Majorana masses for the $N_i$, also referred to as ``right-handed neutrinos,'' for obvious reasons. I choose, without loss of generality, their mass matrix to be diagonal. After electroweak symmetry breaking, ${\cal L}_{\nu}$ will describe, aside from all other SM degrees of freedom, six neutral massive Weyl fermions --- six neutrinos.

Experiments offer some guidance regarding the values of $\lambda_{\alpha i}$ and $M_i$. We are sure of the existence of three neutrinos with masses in the sub-eV range. Furthermore, these three neutrinos are mostly active, {\it i.e.,}\/ they participate in neutral current and, with the charged leptons, charged current interactions. In other words, the neutrinos observed so far have all the interacting properties of those contained in the lepton doublets $L^{\alpha}$.  

Two limiting cases of Eq.~(\ref{eq:seesaw}) easily satisfy all experimental data. If $M_i=0,~\forall i$, the six neutrinos ``fuse'' into three Dirac fermions with a mass matrix given by $m_{\nu}=\lambda v$, where $v$ is the vacuum expectation value of the Higgs doublet. On the other hand, if, $M_i\gg \lambda_{\alpha i} v,~\forall \alpha,i$, the six neutrinos ``split up'' into three light mostly active neutrinos with a Majorana mass matrix whose elements are given by $m^{\alpha\beta}_{\nu}=\sum_i\lambda^{\alpha i}M_i^{-1}\lambda^{\beta i} v^2$, and three heavy, mostly sterile neutrinos \cite{foot2} with Majorana masses $M_i$. The latter limit is the renowned ``seesaw mechanism'' \cite{seesaw}.

What do we know about the magnitude of the seesaw scale $M$? There are several {\sl theoretical} arguments in favor of a very large $M\gg v\sim 100$~GeV. If $M\sim 10^{14}$~GeV, for example, neutrino masses are $O(10^{-1}~\rm eV)$ for $\lambda=O(1)$, and one can plausibly argue that $M$ is related to the grand unification scale, $M_{\rm GUT}$, the energy associated to the unification of the running SM gauge couplings. Thermal leptogenesis \cite{leptogenesis} points to $M\gtrsim 10^{10}$~GeV \cite{leptogenesis_recent}, while one can argue that the seesaw scale may be related to the physics responsible for stabilizing the electroweak scale, such that $M\sim 1-10$~TeV \cite{TeV_seesaw}.   

The following points are worth emphasizing: (i) we know very little about $M$ \cite{strumia_sterile,inprogress} other than $M\in[0,M_{\rm max}]$, where $M_{\rm max}\sim 10^{16}$~GeV from neutrino data and the requirement that physics at the seesaw scale is perturbative \cite{willen}, (ii) small values of $M$ are natural in the sense that $M\to 0$ enhances the symmetry of ${\cal L}_{\nu}$, {\it i.e.,} $U(1)_{B-L}$ is exact in the SM augmented by only Dirac neutrino masses. In summary, there is nothing wrong with $M\ll v$ both experimentally and theoretically  \cite{low_idea,low_model}.

On the other hand, neutrino data may have revealed more than three massive, mostly active neutrino states. Since 1995 \cite{lsnd_first}, the Liquid Scintillator Neutrino Detector (LSND) collaboration has been reporting evidence for a $\bar{\nu}_e$ flux thirty meters away  from a source of neutrinos produced in $\pi^+\to\mu^++\nu_{\mu}$ decay in flight and the subsequent decay, at rest, of the daughter $\mu^+\to\bar{\nu}_{\mu}+e^++\nu_e$. Currently, the evidence is statistically very significant \cite{lsnd}. The unexpected $\bar{\nu}_e$ candidates can be explained if there is a small average probability $P_{\mu e}=(0.26\pm 0.08)\%$ for a neutrino produced as a $\bar{\nu}_{\mu}$ to be detected as a $\bar{\nu}_e$ \cite{lsnd}.
 
 This LSND anomaly, however, is looked upon with suspicion by the majority of the theoretical and experimental neutrino community, for a few reasons. One is that it is yet to be verified by another experiment. This criticism is currently being addressed by the MiniBooNE experiment at Fermilab \cite{miniboone}, and definitive confirmation or refutation of the LSND claims will be obtained in the near future (Fall 2005?). Another is that there is no completely satisfactory solution to the LSND anomaly. I'll elaborate a little more on this point in the next paragraphs.
 
 If interpreted as evidence for neutrino masses and mixing, the LSND data (combined with several negative searches for neutrino oscillations at similar values of $L/E$) require a value of $\Delta m^2$ much larger than the one required to fit the rest of the neutrino data. Therefore, in order to explain the LSND anomaly together with the rest of the neutrino data, at least three mass-squared differences are required, and hence at least four neutrino states with masses in the several eV range or lower exist. LEP measurements of the invisible Z-width \cite{PDG}, require the extra neutrinos to be sterile. Other solutions to the LSND anomaly have yet to fare better than mass-induced flavor change \cite{3+1}, and will be ignored henceforth.
 
 The introduction of extra neutrinos does not guarantee that all data can be properly understood. Indeed, detailed analyses reveal that ``2+2 mass schemes'' are safely ruled out, while ``3+1 mass schemes'' are marginally allowed \cite{3+1,recent_fit,3+2}. The addition of more sterile neutrinos allows for a better fit, and ``3+2'' schemes have been explored \cite{3+2}. If one accepts light sterile neutrinos as the solution to the LSND anomaly, there remains the theoretical complaint that these are ``uncalled for'' and ``unnatural.'' Indeed, most models for light sterile neutrinos (for a brief discussion and many  references, see \cite{strumia_sterile}) are rather cumbersome or contrived, and not particularly predictive.
 
 In this letter,  I wish to point out that the LSND data provides the only {\sl experimental} hint of a higher mass scale in the neutrino sector --- $\sqrt{\Delta m^2_{\rm LSND}}$ --- and pursue whether $M\sim 1-10$~eV (the ``eV-seesaw'' scenario) is consistent with all neutrino data.  It is crucial to appreciate that, once such a relation is imposed, the ``LSND'' mixing angles are determined in terms of active neutrino masses and mixing, rendering the hypothesis raised above predictive and, hence, falsifiable.
    
Define $\lambda_{\alpha i} v\equiv \mu_{\alpha i}$. In the seesaw limit ($\mu\ll M$), 
\begin{equation}
m_{\nu}^{\alpha\beta}=\sum_i\frac{\mu_{\alpha i}\mu_{\beta i}}{M_i}.
\label{mass_seesaw}
\end{equation}
On the other hand, in the basis where the charged lepton mass matrix and the weak interactions are diagonal,
\begin{equation}
m_{\nu}^{\alpha\beta}=\sum_i U_{\alpha i}U_{\beta i} m_i,
\label{mass_mixing}
\end{equation}
where $U_{\alpha i}$ are the elements of the lepton mixing matrix, and $m_i$ are the active neutrino mass eigenvalues. Combining Eqs.~(\ref{mass_seesaw}) and (\ref{mass_mixing}), it is easy to find one solution for $\mu_{\alpha i}$ 
\begin{equation}
\frac{\mu_{\alpha i}}{\sqrt{M_i}}=U_{\alpha i}\sqrt{m_i}~\Rightarrow~\mu_{\alpha i}=U_{\alpha i}\sqrt{M_im_i}. \label{mu_Mm}
\end{equation}
 
 Furthermore, still in the seesaw limit, it is trivial to compute the $|\nu_{\alpha}\rangle$ fraction of the predominantly sterile neutrino state $|M_i\rangle$:
 \begin{eqnarray}
 \langle\nu_{\alpha}|M_i\rangle\equiv\vartheta_{\alpha i}&=&\frac{\mu_{\alpha i}}{M_i}+O\left(\frac{\mu^2}{M^2}\right), \\
 &=&U_{\alpha i}\sqrt{\frac{m_i}{M_i}}+O\left(\frac{m}{M}\right). \label{vartheta}
 \end{eqnarray}  
 For $M_i\sim 1$~eV, and typical values of $m_i$ and $U_{\alpha i}$ (say, $m_i\sim 0.01$~eV and $U_{\alpha i}\sim 1$), $|\vartheta_{i\alpha}|^2\sim~10^{-2}$, which is of the same order as the mixing angles required to explain the LSND anomaly in 3+1 and 3+2 mass-schemes. In order to determine how well eV-seesaw sterile neutrinos fit the LSND anomaly (and under what conditions), it is convenient to discuss in detail some explicit examples. 
 
 In 3+1 schemes, the ``LSND'' mixing angle and mass-squared difference are given by
\begin{eqnarray}
\sin^22\theta_{\rm LSND}=4|U_{e4}|^2|U_{\mu4}|^2, \\
\Delta m^2_{\rm LSND}\sim|\Delta m^2_{i4}|, ~ i=1,2,3.
\end{eqnarray}   
where $\nu_4$ is the fourth lightest neutrino, and assuming that oscillations involving the fifth and sixth states do not contribute in any significant way. In the case of a normal neutrino mass hierarchy \cite{TASI},
$m_3^2\simeq \Delta m^2_{13}\gg m_2^2,m_1^2$, and it is most promising to choose $m_4=M_3$ ($\Delta m^2_{14}=M_3^2$), such that $|U_{\alpha4}|^2=|\vartheta_{\alpha3}|^2$. In this case,  
\begin{equation}
\sin^22\theta_{\rm LSND}=4|U_{e3}|^2|U_{\mu3}|^2\frac{\Delta m^2_{13}}{M_3^2} < 5\times 10^{-4} \label{normal} 
\end{equation}
for $M_3^2=0.92$~eV$^2$ and using three sigma upper bounds on the active oscillation parameters \cite{recent_fit}. Eq.~(\ref{normal}) is too small to explain the LSND anomaly (the best fit value, according to \cite{3+2}, is $\sin^22\theta_{\rm LSND}=3.1\times 10^{-3}$). 

On the other hand, in the case of an inverted mass hierarchy, $m_2^2\sim m_1^2 \sim |\Delta m^2_{13}|$, and one can choose $m_4$ to agree with either $M_1$ or $M_2$ \cite{foot3}. If $m_4=M_2$ ($\Delta m^2_{14}=M_2^2$), 
\begin{eqnarray}
&|U_{e4}|^2\simeq0.020\left(\frac{|U_{e2}|^2}{0.3}\right)\sqrt{\left(\frac{\Delta m^2_{13}}{3\times 10^{-3}~\rm eV^2}\frac{0.92~\rm eV^2}{M^2_2}\right)}; \\
&|U_{\mu4}|^2\simeq0.024\left(\frac{|U_{\mu2}|^2}{0.42}\right)\sqrt{\left(\frac{\Delta m^2_{13}}{3\times 10^{-3}~\rm eV^2}\frac{0.92~\rm eV^2}{M^2_2}\right)},
\end{eqnarray}
in good agreement with 3+1 fits to the LSND anomaly \cite{3+1,recent_fit,3+2} (according to \cite{3+2}, the best fit point is at $|\Delta m^2_{14}|=0.92$~eV$^2$, $|U_{e4}|^2=0.018$, $|U_{\mu 4}|^2=0.042$).   
 
 In a generic eV-seesaw scenario, more sterile neutrinos are expected to contribute to ``LSND oscillations.'' For example, if the active neutrino masses are quasi-degenerate at $m_1^2\sim m_2^2\sim m_3^2=10^{-2}$~eV$^2$, and $M_3=5$~eV, $M_2=1$~eV (and $M_1$ is large enough in order to not to get in the way, say $M_1\gtrsim 10$~eV) one obtains \cite{foot4}  \begin{eqnarray}
 & \Delta m^2_{15}\sim 25~\rm eV^2, \\
 & \Delta m^2_{14}\sim 1~\rm eV^2, \\
 & |U_{e4}|^2\sim 0.02,~|U_{\mu 4}|^2\sim 0.03 \\
 & |U_{e5}|^2\sim 0.001,~|U_{\mu 5}|^2\sim 0.01
 \end{eqnarray} 
 for allowed values of the active oscillation parameters. Such values provide a decent $3+2$ fit to all neutrino data (the best fit value, according to \cite{3+2}, is at $\Delta m^2_{14}=0.92$~eV$^2$, $\Delta m^2_{15}=22$~eV$^2$, $|U_{e4}|^2=0.01$, $|U_{\mu4}|^2=0.04$, $|U_{e5}|^2=0.002$, and $|U_{\mu5}|^2=0.04$).
 
 In summary, for $M\sim 1-10$~eV, typical sterile--active mixing angles are very close to those required to explain the LSND anomaly. Upon closer inspection, it is clear that, for a normal active neutrino mass-hierarchy ($m_3^2\gg m^2_{2,1}$), the induced mixing turns out to be too small. For an inverted hierarchy ($m_3^2\ll m^2_{2,1}$) or quasi-degenerate active neutrinos ($m_3^2\sim m_2^2\sim m_1^2$), good fits can be obtained. Furthermore, in a generic eV-seesaw, ``3+2'' and ``3+3'' schemes are expected for $M_i/M_j\sim 1$ and $|M_i-M_j|\sim M_i$. Mixing angles capable of addressing the LSND anomaly are obtained as long as some of the active neutrino masses-squared are above few~$\times 10^{-3}$~eV$^2$. 
  
  Other oscillation experiments are also potentially quite sensitive to eV-seesaw neutrinos. The absence of flavor transitions at short baseline experiments (see \cite{strumia_sterile} and references therein) places upper bounds on the magnitude of $m_i/M_i$. These are included in the definition of ``fit all neutrino data including those from LSND'' \cite{3+1,recent_fit,3+2}. Indeed, short-baseline data are the reason 3+1 mass-schemes do not provide a very good fit and 3+2 schemes seem finely tuned \cite{footlast}.  The eV-seesaw scenario alleviates some of the theoretical uneasiness surrounding 3+2 fits by offering an explanation not only for why there are several eV sterile neutrinos but also for why active--sterile mixing $\vartheta^2$ is of order $10^{-2}$ or $10^{-3}$, just outside the reach of short baseline experiments. 
  
  Nontrivial constraints are also imposed by the observation of a $\bar{\nu}_e$-flux from SN1987A that roughly agrees with theoretical expectations (see \cite{strumia_sterile} and references therein). This observation disfavors matter-enhanced $\bar{\nu}_e\to\bar{\nu}_s$ transitions. Significant effects are expected for eV-seesaw neutrinos, especially if these have anything to do with the LSND anomaly. It is not clear, however, which region of parameter space is ruled out by 1987A data given the small number of observed neutrino events and large uncertainties related to the properties of SN1987A and theoretical modelling of supernova explosions. It is fair to say that if the eV-seesaw is indeed correct, extraordinary supernova neutrino flavor transition phenomena are expected, and that these may be revealed with the explosion of a nearby galactic supernova. 
  
The eV-seesaw also has interesting implications for other types of neutrino experiments. For small enough values of $M$, the mostly sterile neutrino states contribute to $m^2_{\beta}$, the effective kinematical neutrino mass-squared probed in precision studies of nuclear beta-decay:
\begin{equation}
m^2_{\beta}=\sum_{i=1}^6|U_{ei}|^2m_i^2\simeq\sum^3_{i=1}|U_{ei}|^2m_i^2+\sum^3_{i=1}|U_{ei}|^2m_iM_i,
\label{effective_mb}
\end{equation}
using Eq.~(\ref{vartheta}). The contribution of the mostly sterile neutrinos  (second sum) dominates that of the mostly active ones (first sum). For example, in the $3+2$ scenario sketched above, the largest right-handed neutrino mass $M_1$ contribution to $m_{\beta}^2$ is
\begin{equation}
m_{\beta}^2\simeq 0.7~{\rm eV^2}\left(\frac{|U_{e1}|^2}{0.7}\right)\left(\frac{m_1}{0.1~\rm eV}\right)\left(\frac{M_1}{10~\rm eV}\right). \label{limit_mb}
\end{equation} 
Eq.~(\ref{effective_mb}) does not capture the full effect of heavy neutrino emission in beta-decay for values of $M$ above the several eV range \cite{smirnov_farzan}, such that Eq.~(\ref{limit_mb}) should only be regarded as a qualitative estimate.
  
Searches for neutrinoless double-beta decay ($0\nu\beta\beta$) also serve as probes of light, mostly sterile neutrinos. Naively, the heavier neutrino states would contribute to the effective neutrino mass probed by $0\nu\beta\beta$, $m_{ee}$, as much as the mostly active neutrinos:   
\begin{equation}
m_{ee}=\left|\sum_{i=1}^6U_{ei}^2m_i\right|\sim\left|\sum_{i=1}^3 U_{ei}^2m_i+\sum_{i=1}^3\vartheta_{ei}^2M_i\right|,
\label{mee}
\end{equation}
and $|\vartheta_{ei}^2|M_i\simeq |U_{ei}|^2m_i$, using Eq.~(\ref{vartheta}). In reality, the situation is more subtle. As long as $m_{ee}$ captures the effect of Majorana neutrino exchange for $0\nu\beta\beta$, the contribution of the lighter, mostly active neutrinos and that of the heavier, mostly sterile neutrinos cancel. This is easy to understand. In the $M\to 0$ limit, neutrinos are Dirac fermions, and the rate for $0\nu\beta\beta$ decay vanishes exactly. Furthermore, it is well known that, in the limit $\mu\gg M$ (pseudo-Dirac neutrinos \cite{pseudodirac}), $m_{ee}$, to a good approximation, also vanishes \cite{0nubb_pseudodirac}. In the scenario discussed here, this is still the case for $M\ll 1$~MeV \cite{inprogress}. For large enough values of $M$, however, Eq.~(\ref{mee}) no longer properly captures the contribution of neutrino exchange to $0\nu\beta\beta$, and there is no longer a sterile--active cancellation. For example, in the usual high-energy seesaw (seesaw scale around or above the weak scale), the contribution of the mostly sterile neutrinos is much smaller than that of the mostly active ones, by a factor $O(Q^2/M^2)$, where $Q^2\sim (50)^2$~MeV$^2$, typical of four-momentum transfers in $0\nu\beta\beta$.   
  
 Cosmological probes may provide the most stringent constraints on the eV-seesaw. Data on the primordial light-element abundances and on fluctuations of the cosmic background radiation spectrum, plus studies of the large scale structure of the Universe, {\sl combined with the ``standard models'' of particle physics and cosmology}, virtually rule out an eV seesaw scale and, more generally, any light, sterile neutrino solution to the LSND anomaly (see, for example, \cite{strumia_sterile}, and references therein). Hence, an eV-seesaw implies nonstandard particle physics or cosmology, such as a very low reheat temperature, or the existence of very light (pseudo)scalar fields \cite{nonstandard}.
  
 What are the down-sides to a low-energy seesaw scale? First, any obvious connection of $M$ to other theoretically well-justified high energy scales ($M_{\rm Planck}$, $M_{\rm GUT}$, $v$) is lost. It is however, quite possible that more subtle connections exist via new flavor or gauge symmetries, etc \cite{low_model,inprogress}.
 
 Second, the neutrino Yukawa couplings turn out to be tiny ($\lambda\sim 10^{-12}$ for $M\sim 1$~eV), and one is entitled to argue that Eq.~(\ref{eq:seesaw}) does not ``explain'' why neutrino masses are small. This is similar to saying that the Yukawa interaction $\lambda_eLH^{\dagger}e^c$ does not ``explain'' why the electron mass is much smaller than 100~GeV. Again, small Yukawa couplings and seesaw masses may be indicative of still-to-be-uncovered new physics.
 
Third, a very low seesaw scale ($M\ll 1$~GeV) does not allow ``canonical'' thermal leptogenesis \cite{leptogenesis,leptogenesis_recent}. Eq.~(\ref{eq:seesaw}) does provide, however, new sources of CP-invariance violation, which may have something to do with the baryon asymmetry of the Universe by ``less canonical'' means.
  
In conclusion (and in summary), Eq.~(\ref{eq:seesaw}) provides a very simple mechanism for rendering the neutrinos massive, in agreement with experimental data. Eq.~(\ref{eq:seesaw}) is the most general  renormalizable Lagrangian consistent with gauge invariance and the existence of the $N_i$ fields, and there is no bottom-up hint of what the value of $M$ should be. Indeed, naturalness arguments may be evoked to point out that $M$ should be very small, in the sense that the system enjoys a larger symmetry structure in the limit $M\to0$. While theoretical prejudice points to a seesaw scale at or significantly above the weak scale, it is important to ask whether experiments have anything to say about the issue.  

The LSND anomaly provides the only ``other'' neutrino-mass scale: $\sqrt{\Delta m^2_{\rm LSND}}\sim 1-10$~eV. If the seesaw scale is postulated to be equal to $\sqrt{\Delta m^2_{\rm LSND}}$, the ``LSND mixing angle'' is solely determined in terms of the elements of the MNS matrix, the active neutrino masses, and $\Delta m^2_{\rm LSND}$. It is, hence,  remarkable that a decent fit to all neutrino data is obtained if at least two of the active neutrino masses are around the 0.05~eV level.

The eV-seesaw predicts the existence of six Majorana neutral fermions lighter than several eV. It predicts vanishingly small rates for neutrinoless double-beta decay, ``large'' neutrino mass effects in tritium beta decay, and rich supernova neutrino oscillations. It also requires the existence of new ingredients to the time evolution of the Universe.  


I am happy to thank Bogdan Dobrescu and Alessandro Strumia for enlightening discussions and comments on the manuscript. This work is sponsored in part by DOE grant \# DE-FG02-91ER40684.


\begin{thebibliography}{99}
\bibitem{foot1} For concreteness, I'll assume there are three $N_i$ fields.
\bibitem{foot2} Neutrinos are said to be sterile if they do not participate in neutral and charged current interactions.
\bibitem{foot3} By assuming a very mild singlet neutrino mass hierarchy, one can ``pick'' one $M_i$ over the others.
\bibitem{foot4} These values were obtained by numerically diagonalyzing the $6\times6$ neutrino mass matrix, starting with $\mu_{\alpha i}$ given by Eq.~(\ref{mu_Mm}). $m/M$ corrections to $\vartheta_{\alpha i}$ are under control.  On the other hand, $m/M$ corrections to $\Delta m^2_{ij}$ (which do not affect any of the discussions here) are very significant.
\bibitem{footlast} 3+2 schemes are finely tuned in the sense that the allowed mixing angles and mass-squared differences carefully avoid constraints from all-but-LSND data while the combined effect of the different oscillation frequencies accounts for the LSND signal. Generically, one should expect ``LSND'' oscillations to have also been seen in past $\nu_e$ or $\nu_{\mu}$ disappearance searches.
\end{thebibliography}
 \end{document}